\documentclass[twocolumn,nofootinbib,showpacs,amsmath,amssymb]{revtex4}
\usepackage{amsfonts,amssymb,amscd,amsmath}
\usepackage{graphicx}
\usepackage{subfigure}
\usepackage{bm}
\graphicspath{{figures/}}
\def\gsim{\mathrel{\rlap{\lower4pt\hbox{\hskip1pt$\sim$}}
 \raise1pt\hbox{$>$}}}

 \newcommand\ra{\rangle}
 \newcommand\beq{\begin{equation}}
 
 \newcommand\eeq{\end{equation}}
 \newcommand\beqn{\begin{eqnarray}}
 \newcommand\eeqn{\end{eqnarray}}

\def\fm{\,\mbox{fm}}
\def\GeV{\,\mbox{GeV}}

\def\lsim{\mathrel{\rlap{\lower4pt\hbox{\hskip1pt$\sim$}}
    \raise1pt\hbox{$<$}}}         
\def\gsim{\mathrel{\rlap{\lower4pt\hbox{\hskip1pt$\sim$}}
    \raise1pt\hbox{$>$}}}         

\def\fm{\,\mbox{fm}}
\def\GeV{\,\mbox{GeV}}

\begin{document}
\title{Distinctive features of hadronizing heavy quarks}

\author{B. Z. Kopeliovich$^1$}
\email{boris.kopeliovich@usm.cl}
\author{Jan Nemchik$^{2,3}$}
\email{nemcik@saske.sk}
\author{I. K. Potashnikova$^1$}
\email{irina.potashnikova@usm.cl}
\author{Ivan Schmidt$^1$ }
\email{ivan.schmidt@usm.cl}

\affiliation{$^1$Departamento de F\'{\i}sica,
Universidad T\'ecnica Federico Santa Mar\'{\i}a,\\
Avenida Espa\~na 1680, Valpara\'iso, Chile}
\affiliation{$^2$
Czech Technical University in Prague, FNSPE, B\v rehov\'a 7, 11519
Prague, Czech Republic}
\affiliation{$^3$
Institute of Experimental Physics SAS, Watsonova 47, 04001 Ko\v
sice, Slovakia
}

\begin{abstract}

The color field of a quark, stripped off in a hard reaction, is regenerated
via gluon radiation. The space-time development of a jet is controlled by
the coherence time of gluon radiation, 
which  for heavy quarks  is subject to the dead-cone effect, 
suppressing gluons with small transverse momenta.
As a result, heavy quarks can radiate only a small fraction of the initial energy. This explains the peculiar shape of the measured heavy quark fragmentation function, which strongly peaks at large fractional momenta $z$. The fragmentation length distribution, related to the fragmentation function in a model independent way, turns out to be concentrated at distances
much shorter than the confinement radius. This implies that the mechanisms of heavy quark fragmentation is pure perturbative.

\end{abstract}
\maketitle

\section{Radiative energy loss in vacuum}
\label{eloss}

A parton originated from a hard reaction has to regenerate its color field, which was 
shaken off due to the strong kick acquired at the origin. The latter is characterized by the hard scale of the reaction, $Q^2=q_T^2+m_Q^2$, where $q_T$ and $m_Q$ are the  quark transverse momentum and mass respectively. For the sake of concreteness we consider jets, produced in $e^+e^-$ annihilation, or initiated by a heavy quark produced at the mid rapidity in a high-energy $pp$ collision.

The process of field regeneration is accompanied by gluon radiation,
forming a jet of hadrons. These gluons are carrying a part of the total jet energy, and since
the radiation process has time ordering, it can be treated as energy dissipation, or vacuum energy loss.

The coherence length of gluon radiation by  a heavy quark
of energy $E$ is given by the inverse longitudinal momentum transfer $q_L=(M_{gQ}^2-m_Q^2)/2E$. The quark-gluon invariant mass $M_{gQ}^2=k_{\perp}^2/x(1-x)+m_Q^2/(1-x)$, where $x$ and $k$ are the fractional light-cone (LC) momentum of the gluon and its transverse momentum relative to the jet axis. The Fock components of the quark, $|Q\ra$ and $|gQ\ra$, get out of phase, i.e. become incoherent, on the characteristic distance, called coherence length,
\beq
L^g_c=\frac{2E\,x(1-x)}{k_{\perp}^2+x^2\,m_Q^2},
\label{140}
\eeq
Thus, gluons are radiated sequentially, in accordance to their coherent lengths, rather than burst simultaneously \cite{LP}.
According to Eq.~(\ref{140}) first  
are radiated gluons with small $x$ 
and large transverse momenta, while gluons
with small radiation angles get to mass shell later  \cite{troyan}.

The amount of energy, 
radiated along the quark path length $L$ 
from the hard collision point can be evaluated as \cite{knp,within,similar},
\beq
\Delta E_{rad}(L) =
\int\limits_{\lambda^2}^{Q^2}
dk_{\perp}^2\int\limits_0^1 dx\,\omega\,
\frac{dn_g}{dx\,dk_{\perp}^2}\,
\Theta(L-L^g_c)\,,
\label{130}
\eeq
where $\omega$ is the gluon energy; $Q^2$ is the hard scale characterising the process, e.g. 
the c.m. energy squared $s_{e^+e^-}$ in $e^+e^-$ annihilation, or $p_T^2+m_Q^2$ in hadronic collisions. The bottom limit of $k_{\perp}^2$ integration $\lambda\approx 0.7\GeV$ corresponds to the onset of nonperturbative effects  \cite{kst2,jet-lag}.
The step function $\Theta(L-L^g_c)$
selects only those gluons, whose radiation length $L_c^g<L$.

The spectrum of radiated gluons in (\ref{130}) has the form,
\beq
\frac{dn_g}{dx\,dk_{\perp}^2} =
\frac{2\alpha_s(k_{\perp}^2)}{3\pi\,x}\,
\frac{k_{\perp}^2[1+(1-x)^2]}{[k_{\perp}^2+x^2m_Q^2]^2}\,.
\label{145}
\eeq
This expression shows that gluon
radiation is subject
to the \textit{dead cone effect} \cite{troyan},
which implies that gluons with small
$k_{\perp}^2 < x^2 m_Q^2$ are suppressed.
Consequently,
heavy quarks radiate less energy
than the light ones. Moreover, the suppressed part of the radiation spectrum, small $k_{\perp}^2$, is responsible for gluon radiation at long distances. Therefore, one should expect a faster regeneration of the color field of heavy quarks in comparison with light ones.

\begin{figure}[hbt]
\vspace{-0.2cm}
    \includegraphics[height=8.5cm]{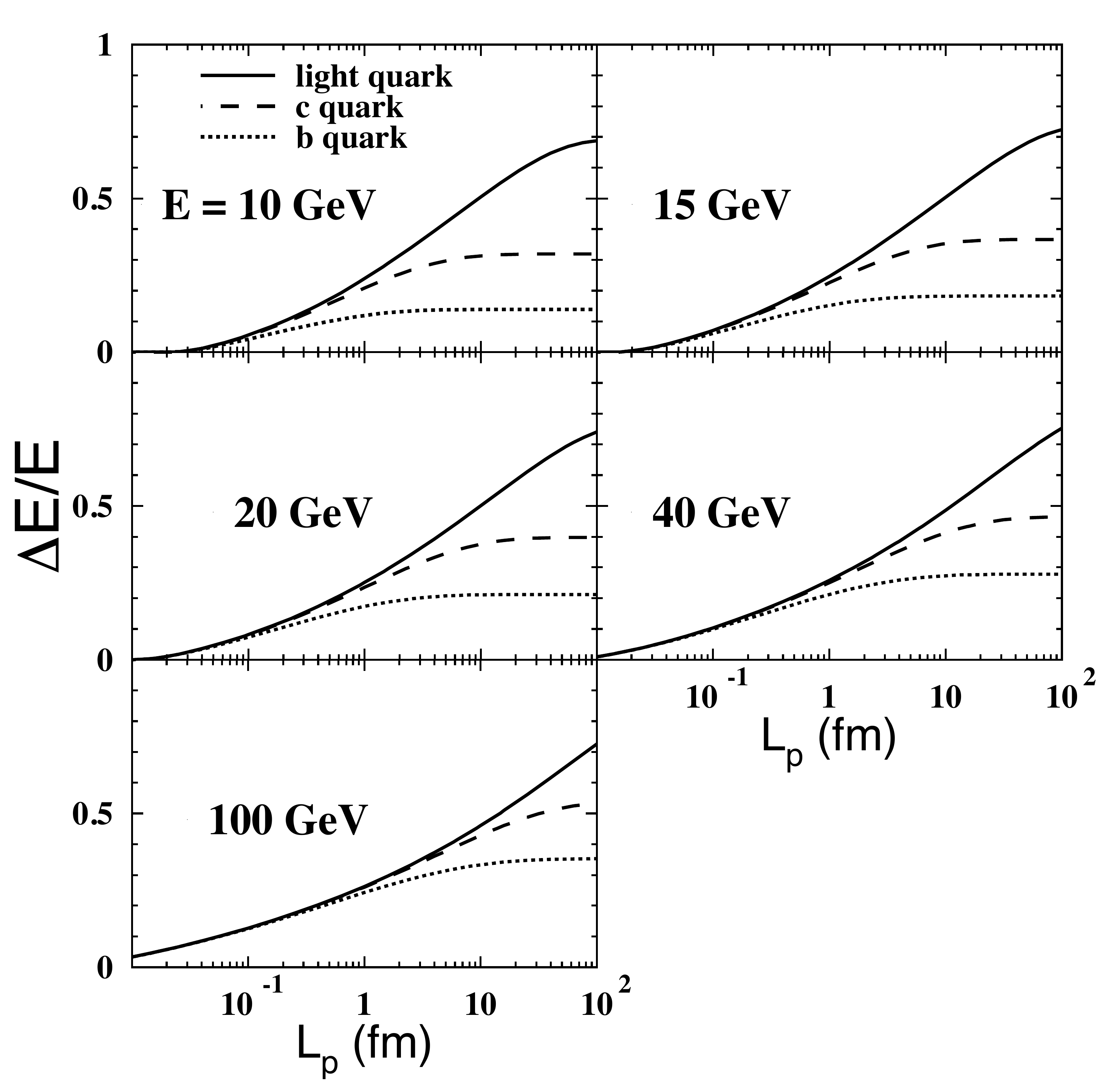}
    \caption{ \label{fig:eloss-qn}
         Fractional radiational  energy loss in vacuum by light, $c$ and $b$ quarks 
         as function of path length for  
         different initial quark energies.}
\end{figure}

Such a behavior can be seen in Fig.~\ref{fig:eloss-qn}, demonstrating 
the $L$-dependence of the radiated fraction of energy 
by different quark species. 
For this calculation we considered the case of the processes with "maximal"
hard scale, like $e^+e^-$ annihilation, or high-$p_T$ production at the mid rapidity,
in the collisions c.m. frame. In this situation one cannot vary independently the jet energy $E$ and the scale $Q^2$, which are strictly related.

\begin{figure}[hbt]
\centering
\includegraphics[width=8.5cm,clip]{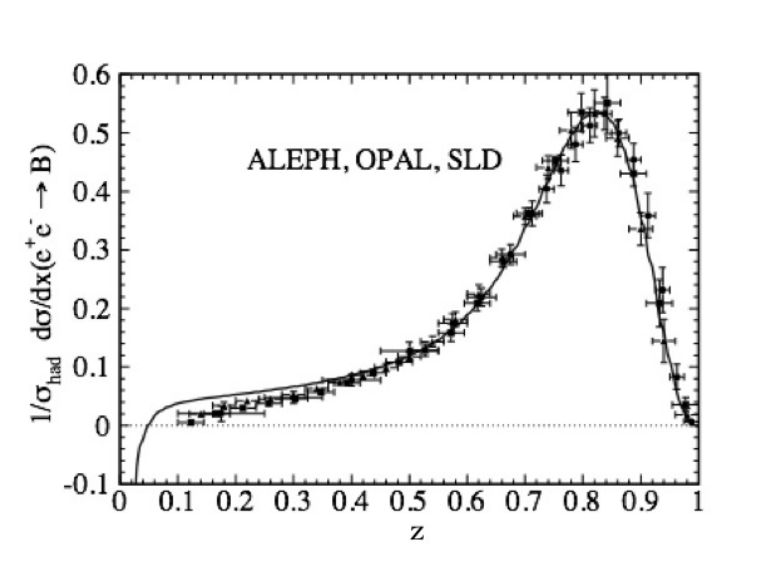}
\vspace{-0.5cm}
\caption{The $b\to B$ fragmentation function, 
         from $e^+e^-$ annihilation. 
         The curve is the DGLAP fit \cite{bottom}.}
\label{fig:ff}
\end{figure}

Remarkably, the results depicted in Fig.~\ref{fig:eloss-qn} demonstrate that in contrast to light quarks, which
keep radiating over a long distance, loosing 
most of the initial energy,  radiation of heavy quarks ceases shortly, and no energy is radiated afterwards, i.e.
the color field of heavy quark is restored promptly.  Fig.~\ref{fig:eloss-qn} also demonstrate that a hadronizing heavy quark loses only a small part of the initial energy,
i.e. the final $D$ or $B$ mesons carry almost 
the whole momentum of the jet. So the $z$ distribution of produced heavy-light mesons should be concentrated at large values of $z$.
This expectation is confirmed with 
the direct measurements of the fragmentation functions
$c\to D$ and $b\to B$
in $e^+e^-$ annihilation \cite{charm,bottom}.
An example of the $b\to B$ fragmentation function depicted in Fig.~\ref{fig:ff}
 shows that it strongly
peaks at $z\sim0.85$. 
A similar behavior was also observed 
for the $c\to D$ fragmentation function \cite{charm}
with the maximum of the distribution at $z\sim0.60\div 0.65$.

On the contrary, the fragmentation functions
of light quarks to light mesons are  known 
to fall with $z$ steadily and steeply from small $z$ up to  $z=1$ \cite{kkp}.

Within perturbative QCD  a quark, which restored its color field and does not radiate gluons anymore,  can be treated as a free colored particle, contradicting the concept of confinement.
However, the nonperturbative effects (color strings) controlling the final stage of hadronization, result in production of only colorless hadrons.

Remarkably,  Fig.~\ref{fig:eloss-qn} demonstrates an universal $L$-dependence of energy loss at short distances for all species of quarks. This happens due to 
the step function $\Theta(L-L^g_c)$ in Eq.~(\ref{130}), which
creates another, even wider dead cone \cite{similar}, 
leading to suppression of gluon radiation with transverse momenta,
\beq
k_{\perp}^2 <
\frac{2 E x (1 - x)}{L} - x^2 m_Q^2\,.
\label{148}
\eeq
This bound is practically independent of quark masses, but relaxes gradually with time,
reaching the magnitude
$k_{\perp}^2\approx x^2 m_Q^2$ at the distance $L \approx
{E (1 - x)}/{x m_Q^2}$,
where the dead cone effect related to the quark mass, takes over.
At longer distances
the heavy and light quarks
start radiating differently.  

\section{Production length of heavy mesons in vacuum}
\begin{figure}[h]
\centering
\includegraphics[width=4.0cm,clip]{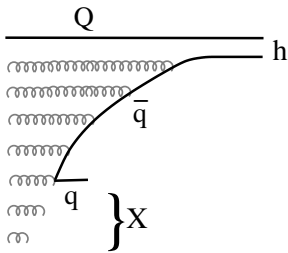} 
\vspace{-0.4cm}
\caption{Redistribution of the energy inside the $Q\bar q$
         dipole. The gluons radiated by $Q$ are absorbed by $\bar q$
         so the dipole energy remains unchanged.  } 
\label{fig:regge-cut}
\end{figure}

Energy loss for gluon radiation by the heavy quark, as well as retarding by nonpertubative
interactions (color string), ceases right after picking up a light antiquark and creation of a colorless heavy-light $Q\bar q$ dipole. The required path length from the origin, we call
the production length, $L_p$. The dipole keeps propagating further, developing the wave function of the detected heavy meson, which might be the ground state of $Q\bar q$, or its excitations. Meanwhile, the $b$ quark inside the dipole might be still virtual radiating gluons and regenerating its color field, but this radiation is to be absorbed by the co-moving $\bar q$ quark, so that the dipole momentum remains unchanged. Such a process of intrinsic radiation is illustrated in Fig.~\ref{fig:regge-cut}.
This cartoon describes the process of picking-up and acceleration of the light antiquark. It 
corresponds to the unitarity cut of a $\bar qq$ Reggeon, and the radiated gluons give rise to reggeization of the $\bar qq$ exchange. Thus, the heavy quark, being a constituent of the $Q\bar q$ dipole, keeps losing energy, sharing it with the light quark,  with the same rate as in vacuum.
Тhe fractional LC momentum $\alpha$ of the light quark in the formed meson
is very small, $\alpha\approx m_q/m_Q$, i.e. is about $15\%$
and $5\%$ for $c$ and $b$ quarks, respectively.

In what follows for the sake of concreteness we consider 
mostly production of $B$
mesons. Extension  to $D$ mesons is straightforward. 

As far as 
the rate of radiational vacuum energy loss $dE/dL$ at $L<L_p$ is known, one
can  relate the production length distribution
$W(L_p)$ to the $b\to B$ fragmentation function $D_{b/B}(z)$.
Indeed, the argument $z$ of the fragmentation function is related to the energy/momentum loss, 
\beq
z\equiv\frac{p_+^B}{p_+^b}=
1-\frac{\Delta p_+^b(L_p)}{p_+^b},
\label{153}
\eeq
where the LC momentum $p^b_+=E+\sqrt{E^2+m_b^2}$ and
its reduction $\Delta p_+^b$ are directly related to the quark energy $E$ and energy loss $\Delta E$,
Eq.~(\ref{130}). Correspondingly, $p^B_+=E-\Delta E+\sqrt{(E-\Delta E)^2+m_B^2}$.
In both cases we neglect the transverse momenta, small in comparison with the heavy flavor masses.

With these kinematic relations and the results of the parameter-free perturbative calculation of $\Delta E$, Eq.~(\ref{130}), the $L_p$ distribution can be directly connected to the fragmentation function,
\beq
\frac{dW}{dL_p}={1\over p_+^b}
\left.\frac{\partial \Delta p_+^b}{\partial L}\right|_{L=L_p}
\!D_{b/B}(z)\, ,
\label{155}
\eeq
where the production length distribution
and the fragmentation function are normalized to unity,
$\int_0^\infty dL_p\, dW/dL_p=1$ and  $\int_0^1 dz\,D_{b/B}(z)=1$.
\begin{figure}[h]
\centering
\includegraphics[width=8.5cm,clip]{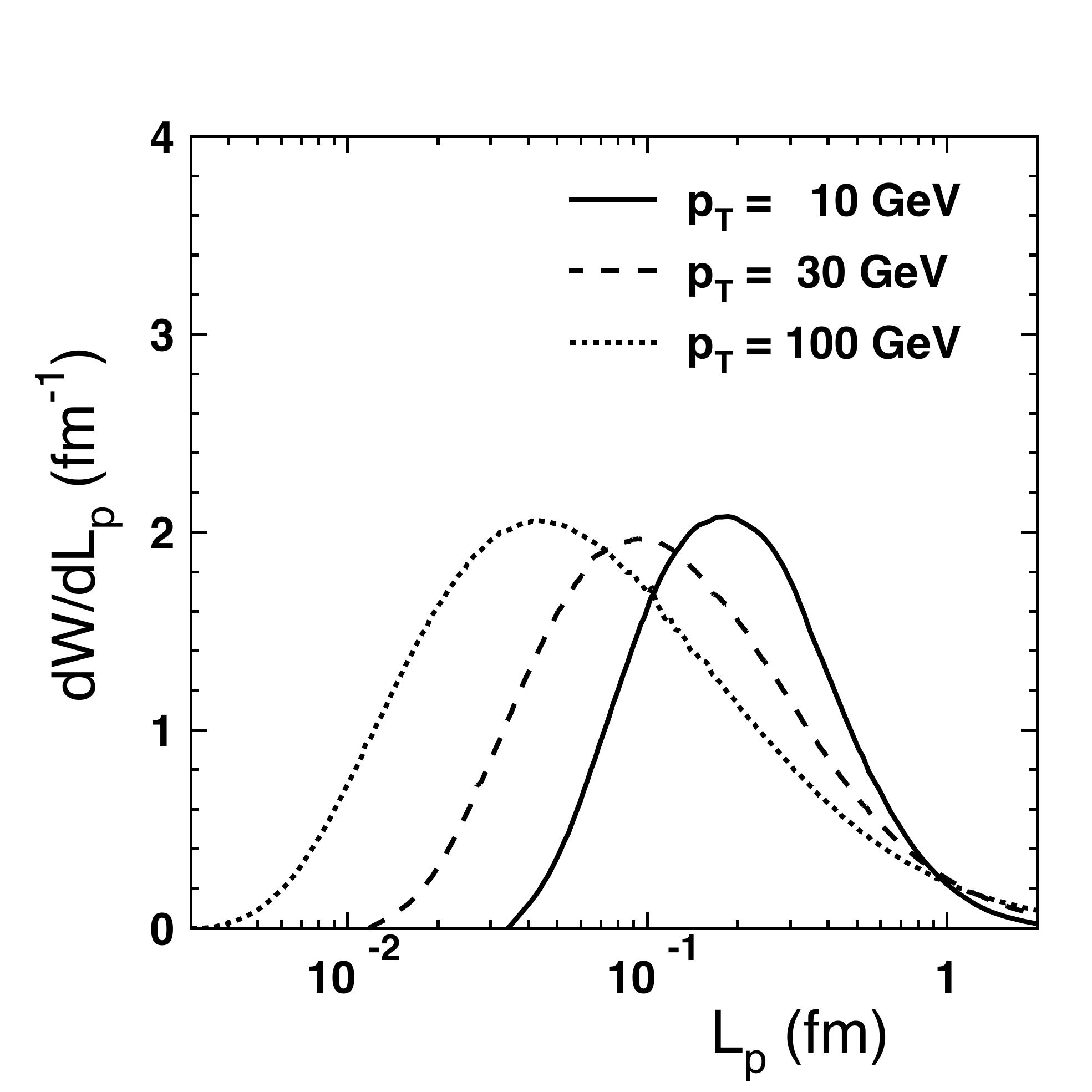}
\caption{The $L_p$-distribution of $B$-mesons produced 
with different $p_T$ in the c.m. frame of $pp$ collisions.}
\label{fig:lp}
\end{figure}

Thus, calculating the shift of fractional momentum 
$\Delta p_+^b(L)$ acquired on the path length $L$ due to gluon radiation,
one can extract the production length distribution directly
from data for $D_{b/B}(z)$. This is a parameter-free and model independent procedure.
Fig.~\ref{fig:lp} shows the $L_p$ dependence of
production length distribution $dW/dL_p$
and clearly demonstrates that
the mean value of $L_p$ shrinks with rising $p_T$,
like it happens for production of high-$p_T$ light hadrons \cite{jet-lag}.
         
Notice that these distances are much smaller than the confinement radius $L_p<<1/\Lambda_{QCD}$, i.e. the fragmentation process so far proceeds in the perturbative regime \cite{pqcd-FF}. Therefore, a large size heavy-light meson cannot be produced, but
only a small-size  colorless $Q\bar q$ dipole, which does not have a certain mass. According to the uncertainty principle the dipole  takes time to form the hadron wave function and acquire a certain mass. This time scale is called formation time or length. The initially produced $b\bar q$ dipole can be expanded over the eigenstates of the mass matrix, i.e. physical states with certain masses ranging from very heavy $M\sim\sqrt{m_b^2+p_T^2}$ down to $m_B$.
The further evolution filters out the states with large relative phase shifts. The longest time takes discrimination between the two lightest hadrons, the ground state $B$ and the first radial excitation $B^\prime$, which concludes the formation process. Correspondingly the full formation path length is,
 \beq
L_f=\frac{2p_T}{m_{B^\prime}^2-m_B^2}.
\label{250}
\eeq
It is controlled by the binding potential. E.g. for oscillatory potential $m_{B^\prime}-m_B=2\omega=0.6\GeV$, so $L_f= 0.06\fm[p_T/1\GeV]$.

\section{Summary}

We demonstrate that the production
of heavy flavored mesons reveal new nontrivial features
in comparison with light hadrons:

\begin{itemize}

\item

The heavy and light quarks originated from high-$p_T$ hadronic  or $e^+e^-$ collisions
radiate very differently.
Heavy quarks are subject to the dead-cone effect and 
radiate a significantly
smaller fraction of the initial energy
regenerating their stripped-off
color field much faster than light ones.

\item
This leads to a specific shape
of the fragmentation functions for
heavy-quark jets.
Differently from light flavors,
the heavy quark fragmentation functions strongly peak
at large fractional momentum, $z\sim 0.60\div 0.65$ and
$z\sim 0.85$ for $c\to D$ and $b\to B$ respectively,
i.e. the produced heavy-light
meson, $B$ or $D$, carry the main fraction of the jet momentum.

\item
The hadronization length distribution is directly related to the measured heavy quark fragmentation function,  Eq.~(\ref{155}), in a model independent way.
The production length distribution depicted in Fig.~\ref{fig:lp} is concentrated 
at extremely short distances, quite shorter than the confinement radius. This implied
a pure perturbative mechanism of heavy quark fragmentation.

\end{itemize}
 
\section{Acknowledgement}
This work was supported in part by grants CONICYT - Chile FONDECYT 1170319 and 1180232, by USM-TH-342 grant, and by CONICYT - Chile PIA/BASAL FB0821.
J.N. work was partially supported by Grants LTC17038 and LTT18002 
of the Ministry of Education, Youth and Sports of the Czech Republic, 
by the project of the European Regional Development Fund 
CZ02.1.01/0.0/0.0/16\_019/0000778, 
and by the Slovak Funding Agency, Grant 2/0007/18.

\end{document}